\documentclass[aps,twocolumn,preprintnumbers,final]{revtex4}

\usepackage{graphicx,amsmath,amssymb,mathrsfs,array,longtable,delarray,verbatim,epsfig}
\graphicspath{{E:\My_papers\Multiplexer}}

\begin{document}

\title{A statistical study of conductance properties in one-dimensional quantum wires, focusing on the 0.7 anomaly}

\author{L. W. Smith$^{1,*}$, H. Al-Taie$^{1,2}$, F. Sfigakis$^{1}$, P. See$^{3}$, A. A. J. Lesage$^{1}$, B. Xu$^{1}$, J. P. Griffiths$^{1}$, H. E. Beere$^{1}$, G. A. C. Jones$^{1}$, D. A. Ritchie${^1}$, M. J. Kelly$^{1,2}$, and C. G. Smith$^{1}$}

\affiliation{
$^{1}$Cavendish Laboratory, Department of Physics, University of Cambridge, J. J. Thomson Avenue, Cambridge, CB3 0HE, United Kingdom\\
$^{2}$Centre for Advanced Photonics and Electronics, Electrical Engineering Division, Department of Engineering, 9 J. J. Thomson Avenue, University of Cambridge, Cambridge CB3 0FA, United Kingdom\\
$^{3}$National Physical Laboratory, Hampton Road, Teddington, Middlesex TW11 0LW, United Kingdom}

\date{\today}
           
\begin{abstract}

The properties of conductance in one-dimensional (1D) quantum wires are statistically investigated using an array of 256 lithographically-identical split gates, fabricated on a GaAs/AlGaAs heterostructure.
All the split gates are measured during a single cooldown under the same conditions.
Electron many-body effects give rise to an anomalous feature in the conductance of a one-dimensional quantum wire, known as the `0.7 structure' (or `0.7 anomaly').
To handle the large data set, a method of automatically estimating the conductance value of the 0.7 structure is developed.
Large differences are observed in the strength and value of the 0.7 structure [from $0.63$ to $0.84\times (2e^2/h)$], despite the constant temperature and identical device design. 
Variations in the 1D potential profile are quantified by estimating the curvature of the barrier in the direction of electron transport, following a saddle-point model.
The 0.7 structure appears to be highly sensitive to the specific confining potential within individual devices.

\end{abstract}

\maketitle

\section{Introduction}

One-dimensional (1D) quantum wires can be defined in a two-dimensional electron gas (2DEG) using split-gate nanostructure devices~\cite{Thornton1986}. 
The conductance of a quantum wire displays plateaus as a function of gate voltage, quantized in units of the conductance quantum $G_0 = 2e^2/h$~\cite{Wharam1988, vanWees1988}, which can be understood within a non-interacting framework. However, an anomalous feature appears near $0.7G_0$--the `0.7 structure' or `0.7 anomaly'~\cite{Thomas1996, Micolich2011}--which occurs as a direct result of electron-electron interactions. This intriguing conductance anomaly continues to inspire efforts to explain its occurrence~\cite{Bauer2013, Iqbal2013}. Many theories have been proposed, including spontaneous spin polarization~\cite{Thomas1996, Wang1996}, quasi-bound state formation and associated Kondo effect~\cite{Cronenwett2002, Meir2002, Rejec2006, Iqbal2013}, and enhanced electron interactions as electrons slow on passing through the 1D channel, due to the potential barrier~\cite{Sloggett2008, Bauer2013, Lunde2009}.

The conductance value of the 0.7 structure ($G_{0.7}$) can vary between $0.5$ and $0.9G_0$~\cite{Thomas1998, Thomas2000, Nuttinck2000, Wirtz2002, Reilly2001, Chiatti2006, Lee2006, Pyshkin2000, Hashimoto2001}, and depends on temperature ($T$), magnetic field ($B$), carrier density in the 1D channel, and device geometry.
The $T$ and $B$ dependence of the 0.7 structure are well established~\cite{Thomas1996}. However, differing results have been reported regarding the dependence on density (summarized in Ref.~\cite{Burke2012}) and upon the length of the 1D channel~\cite{Iqbal2013, Reilly2001}. 
These data were obtained from a variety of 1D devices and confining potentials. This highlights the important role that the confining potential plays in determining $G_{0.7}$~\cite{Reilly2005}.

Due to limited resources available, many previous experimental studies reproduce results on a handful of devices at most.
This prevents statistically significant statements being formulated, and possibly leads to trends being overlooked.
Here, a statistical study of the 0.7 structure is presented, using an array of 256 individual split gates with the same dimensions.
Experiments were performed during a single cooldown, where each split gate was measured separately by means of a multiplexing scheme described in Ref.~\cite{Al-Taie2013}.
Despite the identical device design, fabrication and measurement conditions, large differences exist in the appearance and value of the 0.7 anomaly.
Systematic methods are employed to estimate $G_{0.7}$ and the curvature of the potential barrier in the transport direction.
The value of $G_{0.7}$ appears to be highly sensitive to the specifics of the 1D potential, which differs between even nominally identical split gates.

\begin{figure*}
\includegraphics[width=16cm,height=10cm,keepaspectratio]{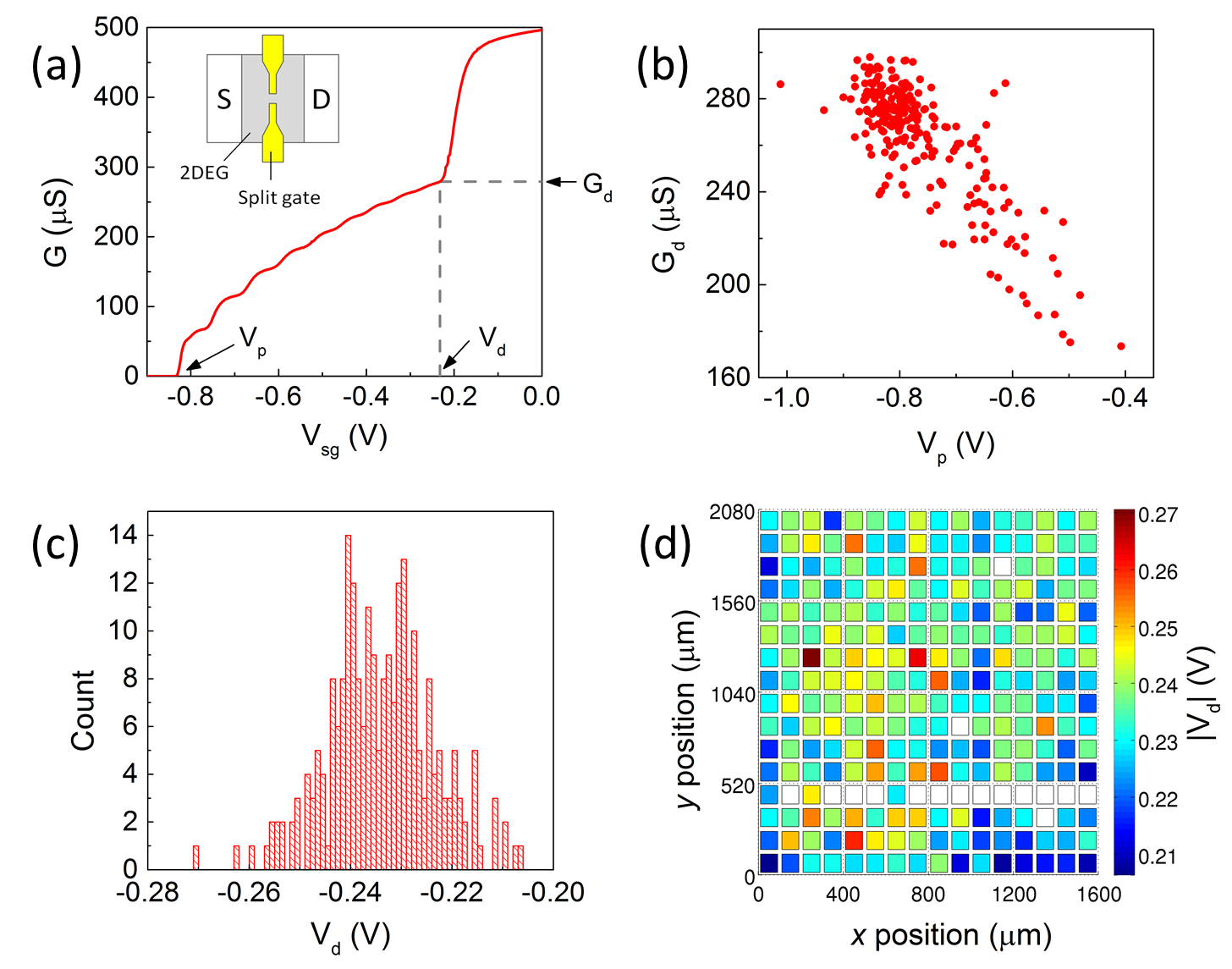}
\caption{\label{Fig1} 
(Color online) (a) Typical trace of conductance $G$ as a function of the voltage applied to the split gate $V_{sg}$, from $V_{sg}$ = 0 to $V_p$. The vertical (horizontal) dashed line indicates $V_d$ ($G_d$). 
The inset shows a schematic diagram of a split gate, where S and D correspond to the source and drain contacts, respectively. 
(b) Scatter plot of $G_d$ against $V_p$ for which $r=-0.81$, indicating a strong negative correlation.
(c) Histogram of $V_d$ from 240 split gates, for a bin size of 1 mV. 
(d) Color scale of $V_d$ as a function of spatial location, where each box represents a split gate in the array. Clear boxes indicate devices which did not define a 1D channel.
The boxes are equally spaced in the $x$ and $y$ directions for convenience, although in reality the split gates have horizontal and vertical pitch lengths of 100 and 130 $\mu$m, respectively.
}
\end{figure*}

The outline of the article is as follows.
First, the methods of fabricating and measuring the sample are described in Sec. II.
We then introduce the properties of the 1D conductance data in Sec. III, highlighting variations which may be related to local fluctuations in the density, thus characterizing the homogeneity of the wafer.
Next, we investigate the 0.7 anomaly and its dependence on properties of 1D conductance in Sec. IV. The method of estimating $G_{0.7}$ is described, to systematically analyze the large data set. 
Since the 0.7 anomaly appears exceedingly sensitive to the 1D potential profile, variations in the potential are quantified by developing a model in Sec. V to estimate the curvature of the barrier in the transport direction, through fitting the data with the transmission probability from a saddle-point model.
Finally, in Sec. VI we acknowledge the role of disorder in giving rise to other anomalous features in conductance at unexpected values, close to $0.5G_0$.

\section{Methods}

The sample was fabricated on a modulation-doped GaAs/AlGaAs high electron mobility transistor (HEMT), in which the 2DEG is formed $90$ nm below the wafer surface. The 2D carrier density ($n_{2D}$) and mobility ($\mu$) of the 2DEG were measured to be $1.7\times10^{11}$ cm$^{-2}$ and $0.94\times10^6$ cm$^2$V$^{-1}$s$^{-1}$, respectively. 
The split gates were arranged in a rectangular array, with pitch lengths of 100 and 130 $\mu$m in the two perpendicular directions. Each split gate was $400$ nm long and $400$ nm wide, defined using electron-beam (\emph{e}-beam) lithography [a schematic diagram of a split gate is shown in the inset of Fig.~\ref{Fig1}(a)]. Two-terminal measurements were performed at $T=1.4$ K, using an ac excitation voltage of 100 $\mu$V at $77$ Hz. Fifteen split gates failed to define a 1D channel, due to damage to one or both arms of the split gate, which is likely to have occurred during fabrication. 
           
\section{Properties of 1D conductance}

Figure~\ref{Fig1}(a) shows a typical conductance trace as a function of the voltage applied to the split gate ($V_{sg}$). Conductance $G$ is plotted from $V_{sg}=0$ to pinch off voltage $V_p$. 
There is an initial drop in $G$ before a quasi-1D channel forms at $V_{sg} = V_d$ (the 1D definition voltage). This is marked by a sudden change in gradient of $G$ as a function of $V_{sg}$. 
The definition conductance ($G_d$), $V_p$ and $V_d$ are indicated by arrows in Fig.~\ref{Fig1}(a).

Figure~\ref{Fig1}(b) shows a scatter plot of $G_d$ against $V_p$ for 240 devices (241 were measured, however, the conductance of one dropped to zero without defining a 1D channel).
The degree of correlation can be quantified using the Pearson product-moment correlation coefficient ($r$), where $r=1$ ($r=-1$) corresponds to a perfect positive (negative) correlation, and $r=0$ corresponds to no correlation.
There is a strong negative correlation between $G_d$ and $V_p$ in Fig.~\ref{Fig1}(b), for which $r=-0.81$.
Since $G$ in these devices is determined by the number of 1D subbands, a higher $G_d$ suggests that there are more 1D subbands in the channel. The subband spacing may therefore be smaller, requiring the channel to be wider on definition. A stronger electric field (more negative voltage) will be required to fully deplete a wider channel, as reflected in Fig.~\ref{Fig1}(b).
No correlations were apparent between $V_d$ and $V_p$ ($r=-0.07$), or $G_d$ and $V_d$ ($r=0.06$).

Figure~\ref{Fig1}(c) shows a histogram of $V_d$, for a bin size of 1 mV. The mean $\bar{V}_d=-0.234$ V and standard deviation $\sigma_{V_d}=10.4$ mV, corresponding to $4.4 \%$ of the mean. Variations in density $n_{2D}$ across the array of split gates may be estimated using $V_d$ from each device  (since $V_d$ is the voltage at which the 2D region beneath the gates is depleted).
By considering the capacitance between the split gate and the 2DEG, $V_d$ is related to $n_{2D}$ by 
\begin{equation}
V_d = \frac{edn_{2D}}{\epsilon}\ ,
\end{equation}
where $e$ is the electronic charge, $d$ is the depth of the 2DEG, and $\epsilon$ is the dielectric constant of the material. 
This equation is valid for gate width $w$ $\gg d$; here $w = 400$ nm and $d = 90$ nm.
For $\bar{V}_d=-0.234$ V, Eg. (1) gives $n_{2D}=1.74\times10^{11}$ cm$^{-2}$ (using $\epsilon \approx 12 \epsilon_0$ for $33 \%$ AlGaAs), and $\sigma_{n_{2D}} = 8\times10^{9}$ cm$^{-2}$. For comparison, conventional Hall bar measurements on two nearby sections of wafer yielded $1.65\times10^{11}$ cm$^{-2}$ and $1.69\times10^{11}$ cm$^{-2}$. 

In this approximation, the capacitance due to the finite density of states in the 2DEG is ignored, since it is small with respect to the geometric capacitance. If it is assumed that changes in density are the only reason for differences in $V_d$, the distribution of $V_d$ is directly proportional to fluctuations in $n_{2D}$ (i.e. standard deviation $\sigma_{V_d} \approx 4\%$ corresponds to the same variation in $n_{2D}$). 

Figure~\ref{Fig1}(d) shows a color scale (color online) of $V_d$ as a function of the position of each device in the array. On the chip, split gates are separated by 100 and 130 $\mu$m in the horizontal and vertical directions, respectively, whereas boxes are equally spaced in Fig.~\ref{Fig1}(d) for convenience. Clear boxes indicate split gates for which $V_d$ could not be determined. 
Plotting $V_d$ as a function of spatial location illustrates how density fluctuations in a HEMT structure can be investigated on a micron scale. 
This technique approximately characterizes the homogeneity of a wafer, since variations on this length scale not shown by conventional Hall bar measurements. 

\section{0.7 structure}

\begin{figure}
\includegraphics[width=12cm,height=14cm,keepaspectratio]{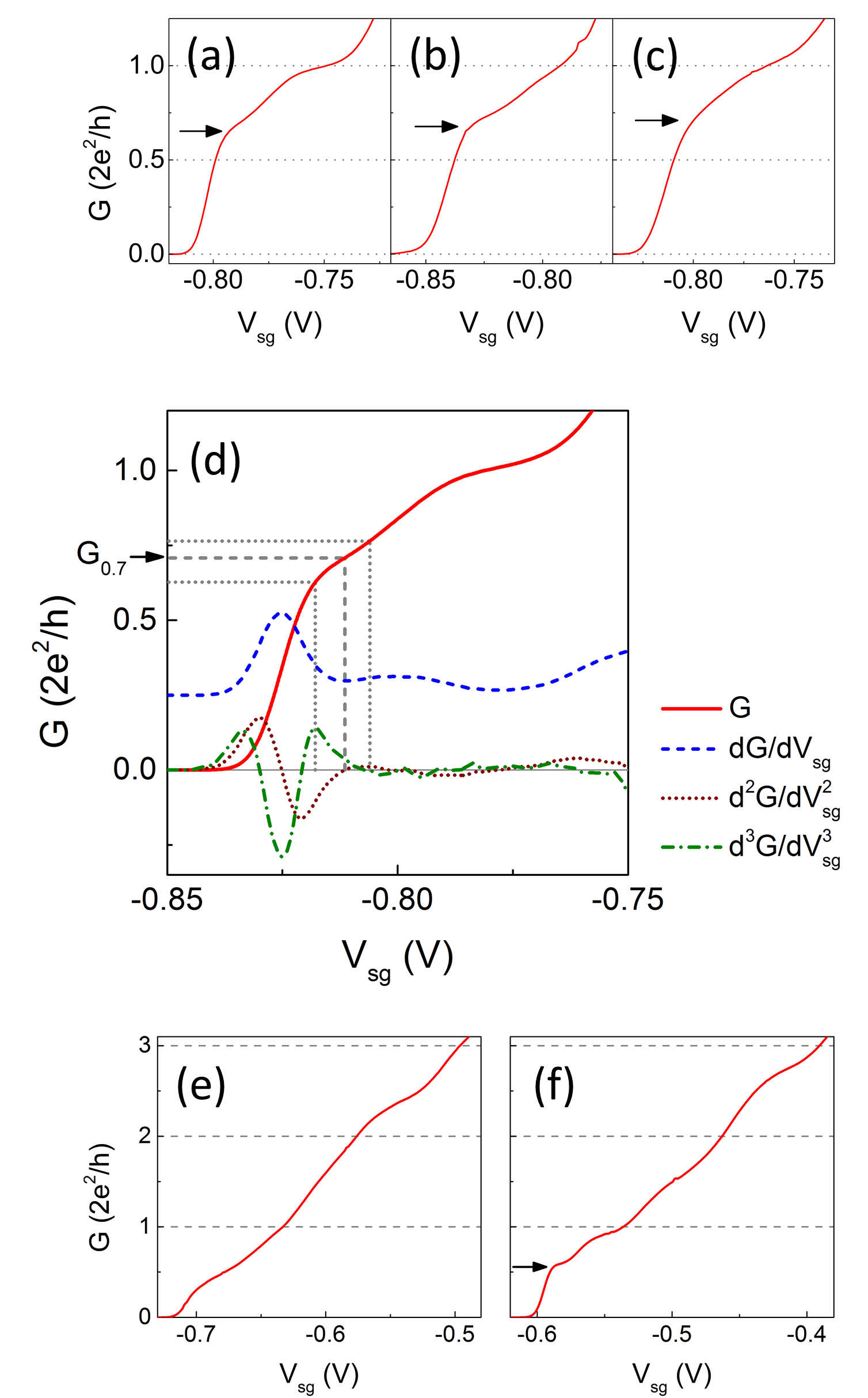}
\caption{\label{Fig2} (Color online) (a)-(c) Conductance against $V_{sg}$ for three nominally identical split gates. Well-developed conductance anomalies occur below $G_0$ in panels (a) and (b), indicated by the arrows. A much weaker shoulder-like feature occurs near $0.7G_0$ in panel (c), marked by the arrow.
(d) Conductance as a function of $V_{sg}$ for a fourth example split gate (solid line). 
First, second and third derivatives $dG/dV_{sg}$, $d^2G/dV_{sg}^2$, and $d^3G/dV_{sg}^3$ are shown by the dashed, dotted, and dash-dotted curves, respectively. The $dG/dV_{sg}$ trace is offset vertically for clarity, and all derivative data are scaled vertically in order to be shown on the plot.
The dashed horizontal line shows $G_{0.7}$, corresponding to the local minimum in $dG/dV_{sg}$.
The upper and lower error bounds of $G_{0.7}$ are shown by the horizontal dotted lines, given by the nearest maxima in $d^2G/dV_{sg}^2$ and $d^3G/dV_{sg}^3$, respectively.
(e), (f) Conductance against $V_{sg}$ for two example split gates showing evidence of disorder. In panel (e), the conductance is no longer quantized in units of $G_0$, while in panel (f) the second plateau is unusually weak and both the second and third plateaus occur below the expected values (the arrow indicates a 0.7 structure).
}
\end{figure}

The conductance data from the array of split gates display a large variation in the appearance of the 0.7 structure.
Figures~\ref{Fig2}(a)-(c) show $G$ as a function of $V_{sg}$ for three example devices, corrected for series resistance $R_s$ (to ensure consistency, $R_s=1/G$ at $V_{sg}=0$).
In Figs.~\ref{Fig2}(a) and 2(b), well-defined structures occur near $0.7G_0$, marked by the arrows.
The feature in Fig.~\ref{Fig2}(b) is particularly pronounced.
A much weaker, shoulder-like structure is shown in Fig.~\ref{Fig2}(c), indicated by the arrow. 

\begin{figure*}
\includegraphics[width=18cm,height=10cm,keepaspectratio]{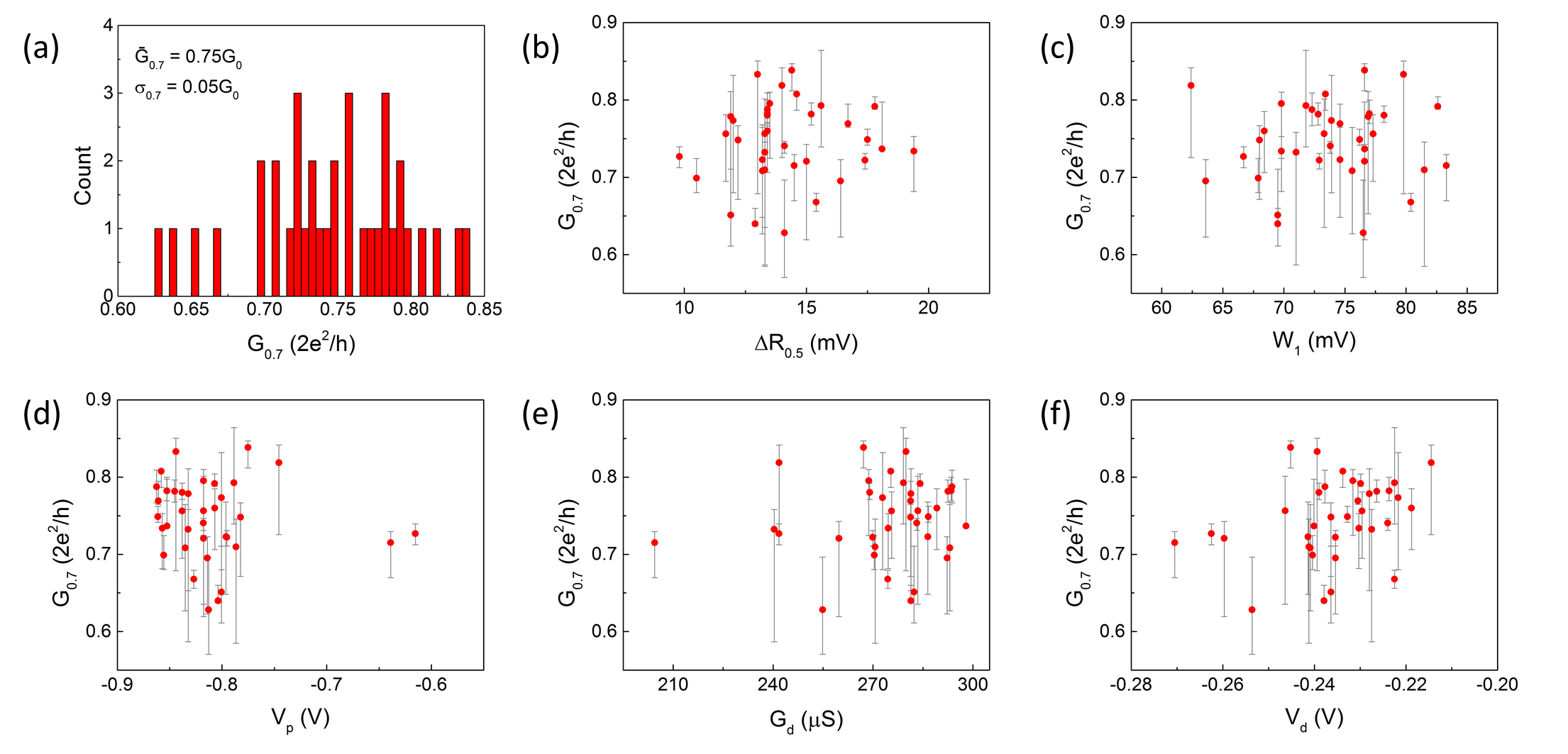}
\caption{\label{Fig3} (Color online) (a) Histogram of $G_{0.7}$, for a bin size of $0.005G_0$. Data are obtained from 36 devices, where $G_{0.7}$ is defined as the local minimum in $dG/dV_{sg}$.
(b)-(f) Scatter plots of $G_{0.7}$ against $\Delta R_{0.5}$, $W_1$, $V_p$, $G_d$ and $V_d$, respectively, where $\Delta R_{0.5} = \Delta V_{sg}$ from $G=0$ to $0.5G_0$, and $W_1=\Delta V_{sg}$ between $G=0.5$ and $1.5G_0$. No strong correlations are apparent; $r=0.07$, $0.07$, $-0.11$, $0.10$ and $0.33$ for (b), (c), (d), (e) and (f), respectively.
Error bounds on the estimate of $G_{0.7}$ are given by the width of the 0.7 anomaly~\cite{noteG07bounds}.
}
\end{figure*}

\subsection{Method of estimating $G_{0.7}$}

A systematic method of estimating $G_{0.7}$ is required to further analyze the data.
There is no stated definition of the conductance value that should be assigned to the 0.7 structure~\cite{note3}; however, it was the location of the flattest part of the feature--most often observed near $0.7G_0$--which led to it being given its particular name~\cite{Thomas2014}. 
Therefore we present data where $G_{0.7}$ is defined as the local minimum in $dG/dV_{sg}$.

Figure~\ref{Fig2}(d) shows $G$ as a function of $V_{sg}$ (solid line) for another example device.
The first, second and third derivatives $dG/dV_{sg}$, $d^2G/dV_{sg}^2$, and $d^3G/dV_{sg}^3$ are shown by the dashed, dotted, and dash-dotted curves, respectively. The $dG/dV_{sg}$ trace is offset vertically for clarity.
The $V_{sg}$ at which $d^2G/dV_{sg}^2=0$ (corresponding to the local minimum in $dG/dV_{sg}$), is shown by the dashed vertical line. This gives our estimate of $G_{0.7}$ (indicated by the arrow). The `width' of the 0.7 plateau is estimated as $\Delta V_{sg}$ between the closest maximum in $d^3G/dV_{sg}^3$ to the left and closest maximum in $d^2G/dV_{sg}^2$ to the right of $d^2G/dV_{sg}^2=0$ (indicated by the vertical dotted lines).
This defines the bounds of our estimate of $G_{0.7}$, shown by the horizontal dotted lines~\cite{noteG07bounds}.

The value of $G_{0.7}$ can be obtained for a limited number of devices using this method, due to variations in shape of the 0.7 structure. If the anomaly is not sufficiently pronounced--for example in Fig.~\ref{Fig2}(c)--$G_{0.7}$ cannot be estimated (since there is no clear minimum in $dG/dV_{sg}$).
However, the resulting benefit is that the data set is reduced to devices for which the conductance characteristics are very similar. Since the strength of the 0.7 structure is related to the relative energy scales within the 1D system, these should therefore be similar for the data remaining. 

In addition, we were careful to discard data which showed evidence of disorder at low $G$ (below $3G_0$), since this may affect $G_{0.7}$.
Various disorder effects were observed. In some instances the quantization of $G$ was significantly affected and no 0.7 structure existed [shown in Fig.~\ref{Fig2}(e)].
In other cases, a 0.7 structure was observed in addition to disorder effects (these effects included missing or weakened plateaus, deviations in plateau values from multiples of $G_0$, unusually weak quantization, and resonant features ranging from Coulomb blockade~\cite{Liang1997} to phase-coherent resonances~\cite{Kirczenow1989}).
Figure~\ref{Fig2}(f) shows $G$ as a function of $V_{sg}$ for a device in which the second and third plateaus occur below the expected values, and the second plateau is weak. The arrow indicates a strong 0.7 structure. 
Such data were discarded because we cannot rule out disorder affecting $G_{0.7}$.

After discarding data which showed evidence of disorder, data from 98 split gates remained ($41\%$ of the 241 measured).
An estimate of $G_{0.7}$ is obtained for 36 of these 98 devices~\cite{07note}.  
This highlights a key benefit of our multiplexing technique: By measuring many devices, we can discard $85\%$ of the data and still retain a data set sufficient for statistical analysis (36 is currently the largest number of devices for which $G_{0.7}$ has been estimated from measurements in a single cooldown).
We expect a lower rejection ratio for a sample fabricated on a wafer with higher mobility.

While we have have attempted to remove the effect of disorder, thermal broadening of energy levels may mask other disorder effects ($T=1.4$ K). 
Measurements were performed at this $T$ in order to observe a well-defined 0.7 anomaly, since we anticipate any statistical (anti)correlation between $G_{0.7}$ and other parameters to be most clear at $T$ for which the 0.7 anomaly is strongest.

\subsection{Dependence of $G_{0.7}$ on properties of 1D conductance}

Figure~\ref{Fig3}(a) shows a histogram of the $36$ counts of $G_{0.7}$ (for a bin size of $0.005G_0$), in which $G_{0.7}$ ranges from $\approx 0.63$ to $0.84G_0$. 
The mean $\bar{G}_{0.7} =0.75G_0$, and standard deviation $\sigma_{0.7} = 0.05G_0$.
The spread of $G_{0.7}$ is within that which has been reported previously~\cite{Burke2012}.
However, Fig.~\ref{Fig3}(a) represents data from devices with a geometrically identical design, which have undergone exactly the same fabrication process, and were measured during a single cooldown at a constant $T$.
Therefore the difference of more than $0.2G_0$ in $G_{0.7}$ is quite remarkable.

Even though each split gate is patterned with a geometrically-identical design, the shape of the 1D potential profile may vary from device to device. This occurs for a number of reasons including fluctuations in $n_{2D}$ (standard deviation of $n_{2D}$ is estimated to be $\approx 4\%$ of the mean, Sec. III), and/or the existence of impurities close to the 1D channel.
The differences in $G_{0.7}$ suggest it is highly dependent on the shape of the potential profile, and minor variations thereof.

In Figs.~\ref{Fig3}(b) to (f), $G_{0.7}$ is plotted against various properties of the 1D conductance trace.
Specifically, Figs.~\ref{Fig3}(b), (c), (d), (e), and (f) show $G_{0.7}$ against $\Delta R_{0.5}$, $W_1$, $V_p$, $G_d$, and $V_d$, respectively, where $\Delta R_{0.5}=\Delta V_{sg}$ from $G=0$ to $0.5G_0$ (corresponding to the steepness of the initial rise in $G$ towards the 0.7 anomaly), and $W_1 = \Delta V_{sg}$ between $G = 0.5$ and $1.5G_0$ (estimating the width of the first conductance plateau).

These properties of the 1D conductance trace are determined by or reflect physical conditions of the system. For example, $V_p$ indicates the strength of the electrostatic field at pinch off; this field is weaker for values of $V_p$ closer to zero, such that the confinement potential is generally shallower. Lower electron densities also often result in $V_p$ closer to zero.
As discussed in Sec. III, $G_d$ and $V_d$ depend on the initial number of 1D subbands in the 1D channel, and fluctuations in $n_{2D}$, respectively.
Additionally, the length of the conductance plateaus depends on the 1D subband spacing, and steepness of the transitions between plateaus depends on the length of the potential barrier in the transport direction (discussed in Sec. V).

A relationship between $G_{0.7}$ and any of these properties may illuminate physical conditions which govern $G_{0.7}$. 
However, no correlations are apparent in Figs.~\ref{Fig3}(b)-(f); [$r=0.07$, $0.07$, $-0.11$, $0.10$ and $0.33$, for 3(b), 3(c), 3(d), 3(e), and 3(f), respectively].
Correlations are perhaps hidden because although the properties of conductance may be primarily related to a particular parameter, they are also subject to other influences.
These data illustrate that $G_{0.7}$ is governed by a combination of conditions and is highly sensitive to the specific potential profile within each device.

\begin{figure}
\includegraphics[width=16cm,height=11cm,keepaspectratio]{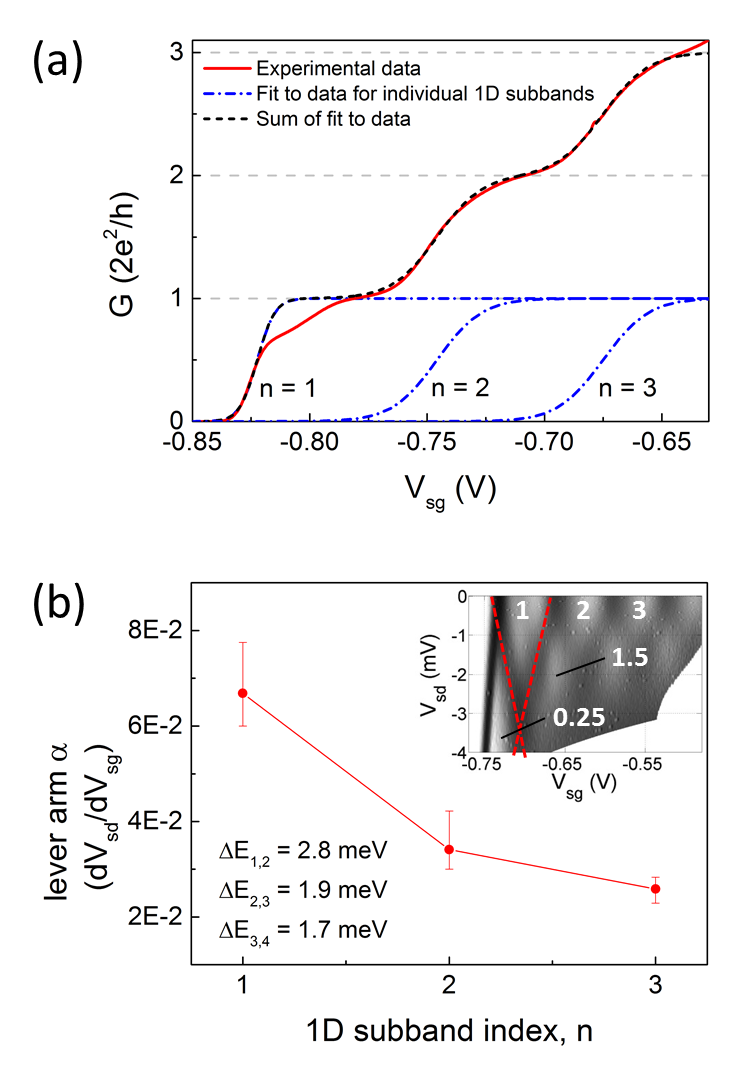}
\caption{\label{Fig4} 
(Color online) (a) The solid line shows the measured $G$ as a function of $V_{sg}$ from an example device [same as Fig. 2(d)]. Dot-dashed lines show a fit to the data ($G_n$) for individual subbands $n=1$, 2, and 3, using a modified saddle-point model~\cite{Buttiker1990}. The points at which $G_n=0.5$ are aligned with the corresponding $G = (n-0.5)G_0$ values on the experimental data, and the dashed line shows $\sum_n G_n$.
(b) Lever arm $\alpha$ as a function of 1D subband index $n$. The inset shows a grayscale diagram of the transconductance $dG/dV_{sg}$ as a function of $V_{sg}$ and $V_{sd}$. The dark (white) regions correspond to high (low) transconductance. The conductance values of low transconductance regions are labeled in units of $2e^2/h$, and $\Delta E_{1,2}$ is given by the point at which the two dashed lines cross.
}
\end{figure}

\begin{figure*}

\includegraphics[width=18cm,height=12cm,keepaspectratio]{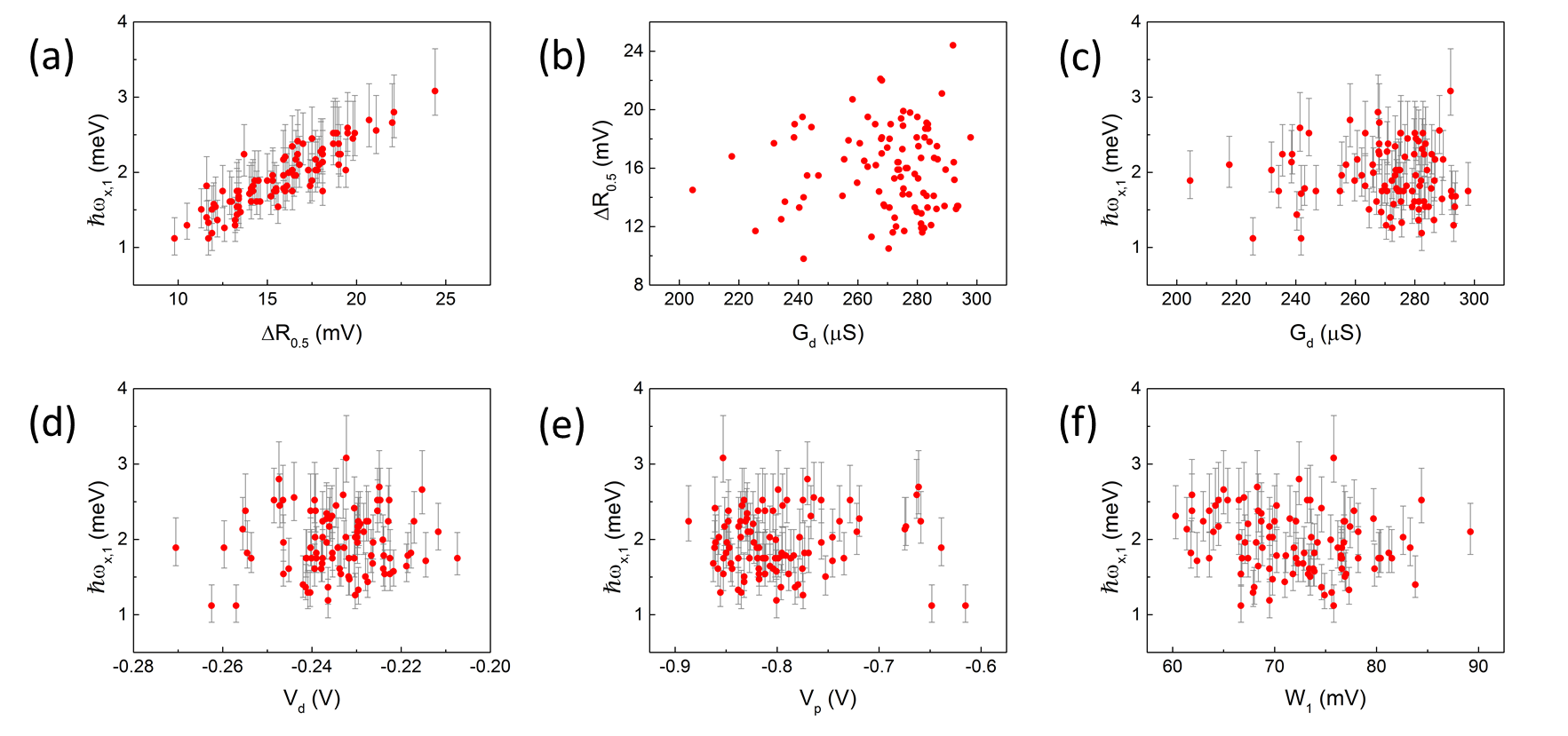}
\caption{\label{Fig5}
(Color online) (a) Scatter plot of $\hbar\omega_{x,1}$ against $\Delta R_{0.5}$, where $\Delta R_{0.5} = \Delta V_{sg}$ from $G=0$ to $0.5G_0$. Error bounds are obtained by finding $\hbar\omega_{x,1}$ for the upper and lower estimates of $\alpha_1$. The strong correlation illustrates the accuracy of the fit to the experimental data.
(b), (c) Scatter plots of $\Delta R_{0.5}$ against $G_d$, and $\hbar\omega_{x,1}$ against $G_d$, respectively. Panel (b) shows a reasonably distinct diagonal cutoff such that there are no data points in the upper-left section of the plot. This is weakly reflected in panel (c).
(d)-(f) Scatter plots of $\hbar\omega_{x,1}$ against $V_d$, $V_p$ and $W_1$, respectively, where $W_1=\Delta V_{sg}$ between $G=0.5$ and $1.5G_0$.
}
\end{figure*}

\section{Quantifying the 1D confining potential}

To quantify the conditions of confinement within each device one can measure the subband spacing using dc bias spectroscopy~\cite{Patel1991}. This type of individual characterization is time consuming, and an automated routine of extracting information from the conductance trace is preferred, because of large data set. 
We therefore fit the data with a transmission probability based on the saddle-point model~\cite{Buttiker1990} in order to estimate the harmonic oscillator energy $E_x =\hbar\omega_x$, which describes the curvature of the potential barrier in the transport direction.

\subsection{Model}

Figure~\ref{Fig4}(a) shows $G$ as a function of $V_{sg}$ (solid line) for an example device. 
The dashed line shows a fit to the data for the transmission probability $T_n=(1+e^{-\pi\epsilon_n})^{-1}$, where $n$ is the 1D subband index, $\epsilon_n = 2[E-E_n-V_0]/\hbar\omega_{x,n}$, $V_0$ is the potential at the center of the 1D channel, and $E_n$ is the energy of the bottom of the $n^{th}$ 1D subband (relative to $V_0$). 
We deviate from a strict saddle-point model which assumes 1D subbands are equally spaced, since this is not the case for real devices.
Additionally, we use a subband-dependent $\hbar\omega_{x,n}$ to achieve a better fit to the data. 

The conductance is calculated independently for $n = 1$, $2$, and $3$ using 
\begin{equation}
G_n = G_0 \int\;dE\; \left(-\frac{\partial f}{\partial E}\right)\;T_n\ ,
\end{equation}
where $f$ is the Fermi-Dirac distribution $f=(1+e^{(E-\mu)/k_BT})^{-1}$.
The conductance increases by $2e^2/h$ when chemical potential $\mu$ rises above the bottom of the 1D subband. We choose a reference frame in which each subband edge is initially at $\mu = 0$, i.e., $V_0$ and $E_n = 0$. The integration is performed between limits $\pm50k_BT$, for $T = 1.4$ K.

The $G_n$ for each subband is then individually scaled by $\alpha_n^{-1}$, where $\alpha$ is a lever arm relating $E$ and $V_{sg}$ obtained from dc bias spectroscopy measurements. We follow the method of estimating $\alpha$ described in the supplementary material of Ref.~\cite{Srinivasan2013}; $\Delta E = \alpha e\Delta V_{sg}$, and $\alpha = \partial V_{sd}/\partial V_{sg}$.
Figure~\ref{Fig4}(b) shows the average values of $\alpha_n$ for each subband from dc bias measurements on four split gates. The error bars are the maximum and minimum estimates of $\alpha_n$.
A grayscale diagram of transconductance $dG/dV_{sg}$ as a function of $V_{sd}$ and $V_{sg}$ from one of the devices is shown as an inset to Fig.~\ref{Fig4}(b). The dark (white) regions correspond to high (low) transconductance, and conductance values of low transconductance regions are labeled in units of $2e^2/h$.
The data are corrected for series resistance (also following the method described in the supplementary material of Ref.~\cite{Srinivasan2013}). 

To automate the fitting routine, $G_n$ for each split gate is scaled by the average $\alpha_n$. For simplicity, we also use a constant $\alpha_n$ for each subband, although in reality it varies with $V_{sg}$.
We believe the use of an average $\alpha_n$ to be the most significant source of error in estimating $\hbar\omega_{x,n}$.

The 1D subband spacings were also obtained from the dc bias data. The average $\Delta E_{n,n+1} = 2.8, 1.9$ and $1.7$ meV, for $n = 1, 2$ and $3$, respectively. 
Since this measurement was performed at $T = 1.4$ K, no feature appears near $0.85 G_0$ at small dc bias \cite{Patel1991, Kristensen2000}. Therefore, $\Delta E_{1,2}$ was estimated as the crossing point of the tranconductance peaks separating the $0.25$ and $G_0$ regions, and the $G_0$ and $1.5 G_0$ regions. 
This is illustrated by the dashed lines on the grayscale [inset, Fig.~\ref{Fig4}(b)]. 

After scaling, the points at which $G_n=0.5 G_0$ are aligned with the corresponding points on the experimental data, i.e., 0.5, 1.5, and $2.5 G_0$ for subbands $n = 1$, $2$ and $3$, respectively.
A fitting routine is used to find the minimum difference squared between the experimental and calculated conductances. For $n=2$ ($3$) the fit was performed between $G=1.01$ ($G=2.01$) and $2 G_0$ ($3 G_0$), with fitting parameter $\hbar\omega_{x,2}$ ($\hbar\omega_{x,3}$). For $n=1$, the fit was performed on the lower half of the riser to the first plateau (from $0.01$ to $0.5 G_0$), to avoid the 0.7 structure.

Figure~\ref{Fig4}(a) shows $G$ after the fitting has been performed, where dot-dashed lines show $G_n$ for individual subbands. The dashed line shows the sum of these data, which overlays the measured $G$ (solid line) well.
The only fitting parameters used in the model are $\hbar\omega_{x,n}$ for $n=1$, $2$, and $3$. Since this is a non interacting model, there is no 0.7 structure in the fit to the conductance data. 

\begin{figure}
\includegraphics[width=8cm,height=10cm,keepaspectratio]{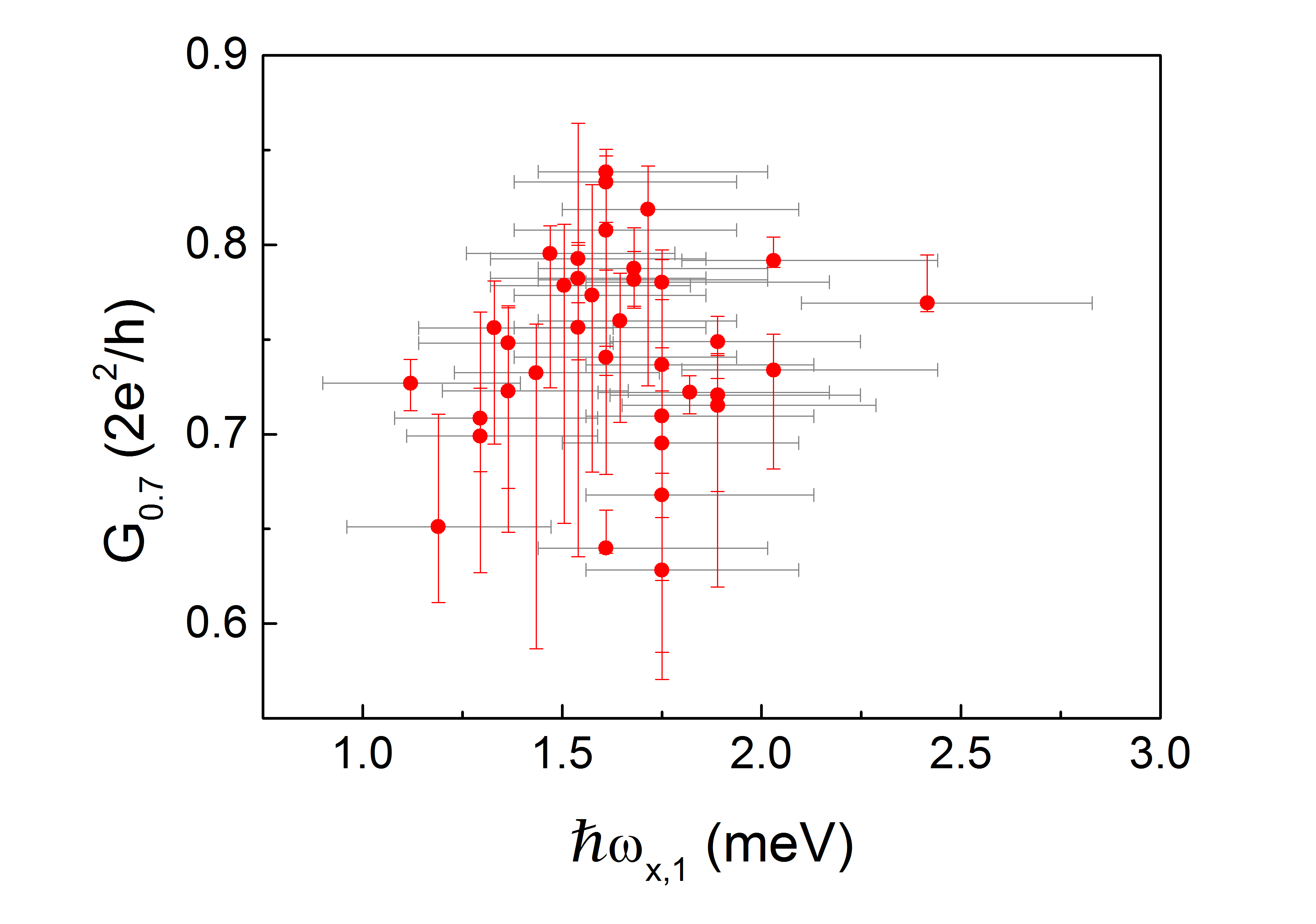}
\caption{\label{Fig6} (Color online) (a) Scatter plot of $G_{0.7}$ against $\hbar\omega_{x,1}$ ($r=0.12$).
Error bounds in $\hbar\omega_{x,1}$ are obtained from the upper and lower estimates of lever arm $\alpha_1$. The bounds on $G_{0.7}$ are related to the width of the conductance anomaly.
}
\end{figure}

\subsection{Dependence of $\hbar\omega_{x,1}$ on properties of 1D conductance}

Using the method described above, $\hbar\omega_{x,n}$ is estimated for the 98 split gates which did not show evidence of disorder below $3G_0$. 
The mean $\hbar\omega_{x,1}=1.95$, $\hbar\omega_{x,2}=1.92$ and $\hbar\omega_{x,3}=1.56$ meV, for which the standard deviations $\sigma_{\hbar\omega_{x,1}}=0.41$, $\sigma_{\hbar\omega_{x,2}}=0.26$ and
$\sigma_{\hbar\omega_{x,3}}=0.22$ meV (corresponding to $21\%$, $14\%$ and $14\%$ of the mean, respectively). 
This highlights the differences that exist in 1D potential from device to device, despite the lithographically identical design. 

The steepness of the initial rise in $G$ towards the 0.7 anomaly is given by $\Delta R_{0.5}=\Delta V_{sg}$ from $G=0$ to $0.5G_0$. Figure~\ref{Fig5}(a) shows a scatter plot of $\hbar\omega_{x,1}$ against $\Delta R_{0.5}$. This is a helpful check of the quality of the fit, since the fitting was performed over this range of $G$. There is a strong degree of correlation (Pearson product-moment correlation $r = 0.9$).
Thus, the steepness of the transition to $G_0$ for the fitted $G$ accurately reflects that of measured data, indicating that the fit is reasonable. Error bounds are given by finding $\hbar\omega_{x,1}$ for the upper and lower values of $\alpha_1$ shown in Fig.~\ref{Fig4}(b).

Figure~\ref{Fig5}(b) shows $\Delta R_{0.5}$ against the 1D definition conductance $G_d$.
While there is no apparent correlation, there seems to be relatively distinct diagonal cutoff above which there are no data points (in the top-left triangular section of the plot). Thus, a sharper initial rise in conductance tends to correspond to a lower $G_d$.
In Fig.~\ref{Fig5}(c), $\hbar\omega_{x,1}$ is plotted against $G_d$. The diagonal cutoff is reflected here, although less distinctly.

Figures~\ref{Fig5}(d),~\ref{Fig5}(e), and~\ref{Fig5}(f) show $\hbar\omega_{x,1}$ against $V_d$, $V_p$, and $W_1$, respectively.
No correlations are apparent between $\hbar\omega_{x,1}$ and these other properties of the 1D conductance trace (no correlations were also evident between $\hbar\omega_{x,2}$ or $\hbar\omega_{x,3}$ and these properties). 
It is possible that trends may be masked by errors in the lever arm $\alpha$. However, the spread in $\hbar\omega_{x,1}$ for most values of $G_d$, $V_d$, $V_p$ and $W_1$ is larger than the estimated error.

\subsection{Dependence of $G_{0.7}$ on $\hbar\omega_x$}

Figure~\ref{Fig6} shows a scatter plot of $G_{0.7}$ against $\hbar\omega_{x,1}$.
For these data $r=0.12$, indicating an exceedingly weak correlation. 
Unfortunately, any correlation which may exist is likely to be masked by errors in the estimate of $\hbar\omega_{x,1}$.
We believe the error in $G_{0.7}$ to be less significant since $G_{0.7}$ is given by a well-defined point (the local minimum in $dG/dV_{sg}$), and bounds in $G_{0.7}$ are instead related to the width of the conductance anomaly~\cite{noteG07bounds}.
The error in $\hbar\omega_{x,1}$ could be reduced by using the correct $\alpha$ for each device. This requires dc bias measurements to be performed for every device.

A specific correlation between $G_{0.7}$ and $\omega_{x}$ is predicted by certain models for the origin of the 0.7 structure. For example, the 1D Kondo effect occurs when electrons are localized within a 1D channel~\cite{Sfigakis2008}. In the 1D Kondo scenario, for large $\omega_x$ the Kondo temperature ($T_K$) is also high~\cite{Meir2002, Rejec2006, Hirose2003}. Thus, an increase in $\omega_x$ ($T_K$) should cause $G_{0.7}$ to increase at a given temperature, since $G \propto [1-(T/T_K)^2]$. However, other models expect the opposite trend. It has been proposed that the 0.7 structure is related to enhanced interactions as the electrons slow down on passing through the 1D barrier~\cite{Sloggett2008, Bauer2013, Lunde2009}. Of these models, only one~\cite{Sloggett2008} studies the high-temperature dependence of $G_{0.7}$ with $\omega_x$.
If the 1D barrier is a saddle-point potential~\cite{Buttiker1990}, the value of $G_{0.7}$ is predicted to decrease as $\omega_x$ increases.
We do not observe a strong enough trend in our data to support one theory above another.

\begin{figure}[t]
\includegraphics[width=8cm,height=10cm,keepaspectratio]{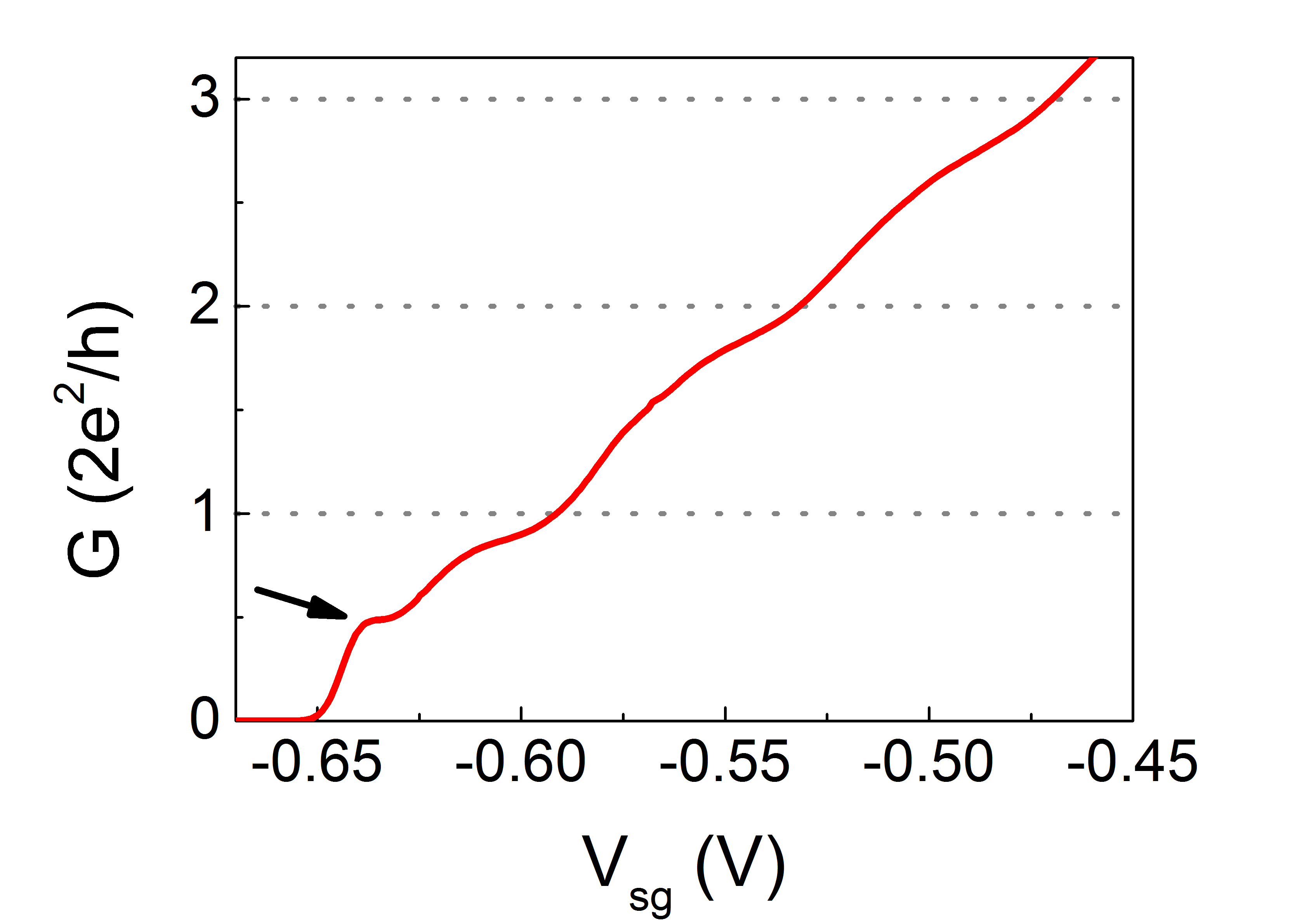}
\caption{\label{Fig7} (Color online) Conductance as a function of $V_{sg}$ for an example split gate, where an anomalous feature appears near $0.5G_0$, indicated by the arrow.
}
\end{figure}

\section{Anomalous conductance features near $G=0.5G_0$}

Conductance anomalies were also observed at $G$ values lower than the range shown in Fig~\ref{Fig3}(a).
Figure~\ref{Fig7} shows $G$ as a function of $V_{sg}$ for a device in which an anomalous feature occurs at $0.5G_0$ (marked by the arrow). 
Of the 241 split gates measured, $\approx 2\%$ showed conductance anomalies at this value; it is therefore unusual to find features at $0.5G_0$ in 1D devices on GaAs/AlGaAs heterostructures for $B=0$ T (the magnetic field at which these measurements were performed).
These data were not included in the analysis because of evidence of disorder; for example, in Fig.~\ref{Fig7} the plateaus do not appear at correct values~\cite{SeriesRes}. The co-existence of disorder effects may be responsible for the lowering of the conductance of the 0.7 structure to $0.5G_0$. 

A statistical measurement makes it possible to distinguish the ``normal'' characteristics from device-specific effects, which may be related to disorder. Devices which display unusual conductance anomaly can be investigated for rare physical phenomena, while very clean devices can be used to identify standard behavior.

The reproducibility of conductance characteristics has been investigated by thermally cycling the sample. 
It was found that many of the split gates which showed evidence of disorder did so on both cooldowns~\cite{Al-Taie2013B}.
The aim of the current article is to compare properties of 1D conductance from a large number of devices on a single cooldown.
The reproducibility of these properties on thermal cycling warrants a further, separate study.

The sample was also illuminated with a light-emitting diode (LED), which increased $n_{2D}$ and $\mu$ to $2.9\times10^{11}$ cm$^{-2}$ and $2.2\times10^6$ cm$^2$V$^{-1}$s$^{-1}$, respectively. This resulted in longer, better-defined plateaus in conductance, since the 1D confining potential becomes stronger and the 1D subband spacing increases. However, many devices showed occasional structure in conductance which had the appearance of resonant transmission through the quantum wire, consistent with an enhancement of resonant effects due to sharper confinement~\cite{Kirczenow1989}.
We did not investigate $G_{0.7}$ or $\hbar\omega_x$ in this case since the estimates are very likely to be affected by the resonances.

\section{Conclusion}

Across an array of nominally identical split gates (measured at a single $T$), significant fluctuations were seen to exist in the 1D potential. These have been quantified by estimating the curvature of barrier in the transport direction ($\hbar\omega_{x}$) for each device.
Large variations were observed in both the appearance of the 0.7 structure 
and the value at which it occurred. The 0.7 structure appears to be extremely sensitive to the specific 1D potential in each device. Measuring many devices has enabled a statistical study to be performed. No correlations were apparent between $G_{0.7}$ and $\hbar\omega_{x}$, or other properties of the 1D conductance trace.

A specific set of physical conditions combine to give a particular conductance trace for a split gate.
With the current analysis, the effect of individual factors influencing the conductance properties cannot be separated.
Thus, parameters which may govern $G_{0.7}$ and $\hbar\omega_x$ have not been identified.
This may become possible by performing dc bias spectroscopy for each device in order to accurately measure 1D subband spacing and lever arm $\alpha$, thereby giving a better estimate of $\hbar\omega_x$.

The confining potential will be affected by fluctuations in the background potential due to the ionized dopants (leading to local density variations) and the existence of impurities (giving rise to disorder effects). We removed data which showed evidence of disorder from the analysis, although a fuller study requires $B$- and $T$-dependent measurements (since disorder effects will be masked at the temperature at which the measurements were performed). We have shown that disorder does give rise to conductance anomaly at unexpected values, e.g., close to $0.5G_0$.
Disorder effects can be reduced by fabricating samples on a wafer with higher electron mobility, or using an undoped heterostructure and electrostatically inducing the 2DEG~\cite{See2012, Harrell1999, Sarkozy2009}.

This work was supported by the Engineering and Physical Sciences Research Council Grant No. EP/I014268/1. The authors thank T.-M. Chen, C. J. B. Ford, I. Farrer, E. T. Owen and K. J. Thomas for useful discussions, and R. D. Hall for \emph{e}-beam exposure.

$*$ Corresponding author. E-mail address: lws22@cam.ac.uk
%\bibliography{aipsamp}% Produces the bibliography via BibTeX.

\begin{thebibliography}{MM}
\bibitem{note3} There is, however, an operational definition: The apparent $0.7$ anomaly should (1) become stronger with increasing temperature, (2) monotonically lower to $0.5G_{0}$ with increasing magnetic field, and (3) should not move up or down in conductance when the conducting 1D channel is laterally shifted left or right with asymmetric voltages on the gates.
\bibitem{noteG07bounds} The bounds of $G_{0.7}$ provide an estimate of the error. However, $G_{0.7}$ is given by a well-defined point (the local minimum in $dG/dV_{sg}$). Also, $\Delta G$ between the upper and lower bounds is very strongly correlated with $\Delta V_{sg}$ between the nearest maximum in $d^2G/dV^2_{sg}$ and $d^3G/dV^3_{sg}$ to the 0.7 anomaly. Therefore, the bounds of $G_{0.7}$ are perhaps better seen as representing the width of the conductance anomaly.
\bibitem{07note} This is not to say that the 0.7 anomaly did not exist in the remainder of devices, but there was no clear minimum in $dG/dV_{sg}$.
\end{thebibliography}

\end{document}